\documentclass[fleqn,usenatbib,useAMS]{mnras}

\usepackage{graphicx}	
\usepackage{amsmath}	
\usepackage{amssymb}	

\usepackage{bm}

\usepackage{multicol}        
\usepackage{bm}		
\usepackage{pdflscape}	
\usepackage{subfig}
\usepackage[export]{adjustbox}

\usepackage{lscape} 
\usepackage{rotating}

\usepackage[T1]{fontenc}
\usepackage{ae,aecompl}
\usepackage{listings}
\usepackage{newtxtext,newtxmath}

\newcommand{\brutus}{\hbox{\sc brutus}}

\newcommand{\dynesty}{\hbox{\sc dynesty}}
\newcommand{\glue}{\hbox{\sc glue}}
\newcommand{\Dataverse}{\hbox{\sc Dataverse}}

\title[North Polar Spur]{Constraining the Distance to the North Polar Spur with Gaia DR2}

\author[]{
Kaustav K. Das,$^{1}$\thanks{E-mail: kaustavn@iitk.ac.in}
Catherine Zucker,$^{2}$\thanks{E-mail: catherine.zucker@cfa.harvard.edu}
Joshua S. Speagle,$^{2}$
Alyssa Goodman,$^{2,5}$
\newauthor
Gregory M. Green,$^{3}$
and Jo\~{a}o Alves$^{4,5}$
\\\\
$^{1}$Department of Physics, Indian Institute of Technology, Kanpur 208016, India\\
$^{2}$Center for Astrophysics $\vert$ Harvard \& Smithsonian, 60 Garden St., Cambridge, MA 02138, USA\\
$^{3}$Max Planck Institute for Astronomy, K{\"o}nigstuhl 17, D-69117 Heidelberg, Germany\\
$^{4}$University of Vienna, Department of Astrophysics, T{\"u}rkenschanzstra{\ss}e 17, 1180 Vienna, Austria\\
$^{5}$Radcliffe Institute for Advanced Study, Harvard University, 10 Garden St, Cambridge, MA 02138
}

\date{Accepted XXX. Received YYY; in original form ZZZ}

\pubyear{2020}

\begin{document}
\label{firstpage}
\pagerange{\pageref{firstpage}--\pageref{lastpage}}
\maketitle
\begin{abstract}
The North Polar Spur (NPS) is one of the largest structures observed in the Milky Way in both
the radio and soft x-rays. While several
predictions have been made regarding the origin  of the NPS, modelling the structure is difficult without precise distance constraints. In this paper, we determine accurate distances to the southern terminus of the NPS and toward latitudes ranging up to 55$^{\circ}$. First, we fit for the distance and extinction to stars toward the NPS using optical and near-infrared photometry and Gaia DR2 astrometry. We model these per-star distance-extinction estimates as being caused by dust screens at unknown distances, which we fit for using a nested sampling algorithm. We then compare the extinction to the Spur derived from our 3D dust modelling with integrated independent measures from XMM-Newton X-ray absorption and
HI column density measures. We find that we can account for nearly 100\% of the total column
density of the NPS as lying within 140 pc for latitudes $>26^{\circ}$ and within 700 pc for latitudes $< 11^{\circ}$. Based on the results, we conclude that the NPS
is not associated with the Galactic Centre or the Fermi bubbles. Instead, it is likely associated,
especially at higher latitudes, with the Sco-Cen association.
 
\end{abstract}




\section{Introduction}

The North Polar Spur (NPS) is a large structure observed at x-ray and radio wavelengths. It is a highly polarised synchrotron source that spans from the Galactic plane to a latitude of $\sim80^{\circ}$, at a longitude of $\sim20^{\circ}$. The NPS is the northern-most part of Loop I, a large circular feature in the radio continuum sky. Several early models for the origin of and distance to the NPS made before the 1980s are summarised in \citet{Salter1983}, which reviews several multi-wavelength observations and theories for its origin. One major theory is that the NPS is part of a local structure (with the closest part being  within $\sim$100 pc) and that it is a part of a supernova remnant. Its prominent X-ray emission, high synchrotron radiation, alignment with starlight polarisation, and alignment with HI filaments at high latitudes (\citet{Heiles2000}, \citet{Heiles1976}, \citet{Axon1976}) all indicated that the NPS is nearby. \citet{deGues1992}, \citet{Wolleben2007} predicted that it is centred at the Sco-Cen OB association. A discussion of other more recent works supporting this theory can be found in the review paper on NPS and Loop-I by \citet{Dickinson2018}. 

There is also evidence that the NPS lies beyond $\sim$100 pc. \citet{Santos2011} detected the NPS via polarised starlight absorption beyond a distance of 100 pc. \citet{Iwan1980} argues that a single supernova model is not sufficient to explain the polarisation absorption and proposed that it lies at a distance of $\sim$400 pc. The similar structure and estimated age as that of the Fermi Bubbles also suggest that the NPS is a Galactic scale structure and located near the Galactic Centre. \citet{Sofue2000} suggested that the NPS is a part of a shock-front of a star-formation activity that took place near the Galactic Centre around 15 Myr ago. Using kinematic distances to the Aquila Rift and Serpens regions, \citet{Sofue2015} argued that the NPS is located beyond 1.1 kpc. \citet{Sun2014} suggested that the Spur is located beyond 2-3 kpc based on the fact that regions below $b<4^{\circ}$ are highly depolarised. However, as \citet{Dickinson2018} argues, the microwave polarisation data and radio maps show that there is little correlation between the Fermi Bubbles and the NPS, as there is no evidence of interaction and the southern portion of the NPS extends far beyond the NPS. Similarly, the \citet{Planck2015} argued that the NPS is not associated with the Galactic Centre based on polarisation maps and geometric constraints. 

A robust constraint on the distance of the NPS has implications for our understanding of the origin and structure of Loop I, the Local Bubble, supernova activities in the solar neighbourhood, and AGN-type outburst activities, as these depend on whether it is a local structure or associated with the galactic centre. In this paper, we obtain distances to the NPS by building upon the methodology described in \citet{Zucker2019} and \citet{Zucker2020}, which utilised a combination of stellar photometry and Gaia DR2 parallax measurements \citep{Lindegren2018} to determine accurate distances to local molecular clouds. We incorporate minor modifications to the model to fit for the distribution of cumulative extinction as a function of distance (extinction ``profiles"), to nineteen fields \citep[from][]{Rosine2016} towards the southern terminus and ten fields towards high latitudes. After fitting for the distance and extinction associated with the NPS, we then compare our inferred extinction at that distance to independent integrated column density measures from the XMM-Newton \citep{Rosine2016} and HI4PI \citep{HI4PI2016} surveys, to determine what fraction of the total integrated emission we can associate with the NPS at different distances.

\section{Data}
\subsection{Photometry and Astrometry}
To obtain estimates of the distance and extinction to stars towards the NPS, we use optical and near-infrared data from the Pan-STARRS1 Survey \citep{Chambers2016} and the Two Micron All Sky Survey \citep[2MASS;][]{Skrutskie2006} as well as astrometry from Gaia DR2 \citep{Gaia2018}.

\subsubsection{Pan-STARRS1}
The Panoramic Survey Telescope and Rapid Response System (Pan-STARRS1) survey images the sky in five photometric filters (\textit{g,r,i,z,y}, spanning from 400-1000 nm), using a 1.8 m telescope and 1.4 Gigapixel camera. The survey spans the sky north of declination of $-30^{\circ}$. A single epoch has a typical 5$\sigma$ point source exposure depth of 22.0, 21.8, 21.5, 20.9 and 19.7 magnitudes for the \textit{g,r,i,z,y} filters respectively (in the AB system). For details on the survey, refer to \citet{Chambers2016}. For this work, we use the `$3\pi$ Steradian Survey', which includes multi-epoch observations carried out over four years for three-quarters of the sky. We include data from the north equatorial pole and use astrometric and photometric calibrations from \citet{magnier2016}. 
\hbadness=11000
 \subsubsection{2MASS}
 The Two Micron All-Sky Survey is an all-sky survey in the infrared. Data are collected using two 1.3 m telescopes in three photometric filters (J, H, and $K_{s}$ spanning from 1-2  $\mu$m). The survey used 7.8 seconds integration time for each pointing in the sky and achieved a typical 10$\sigma$ point source exposure depth of 15.8, 15.1 and 14.3 magnitudes for the J, H, and $K_{s}$ filters respectively (in the Vega system). Further details regarding the 2MASS survey can be found in \citet{Skrutskie2006}. In this work, we use the `high-reliability' catalogue\footnote{Further details in: \url{https://old.ipac.caltech.edu/2mass/releases/allsky/doc/sec2_2.html}} which is devoid of the contamination and uncertainty caused by neighbouring/extended sources.  

\subsubsection{Gaia DR2}
The second release of the Gaia mission provides data for the following astrometric parameters: positions on the sky $(\alpha,\delta)$, parallaxes, and proper motions for over 1.3 billion sources. It provides photometric fluxes in the G, $G_{RP}$, and $G_{BP}$ bands. Further details regarding the data release can be found in \citet{Gaia2018}. We use the astrometric catalogue from \citet{Lindegren2018}, which has a limiting magnitude of 3$\sigma$, and astrometric uncertainty of $\sim{0.04}$ mas for bright stars and $\sim{0.7}$ mas for very faint stars. In this work, we incorporate the astrometric measurements (and their errors) only. 

\subsection{Column Density Estimates}
To determine the fraction of the total integrated column density we can account for at various distances via our 3D dust modelling, we utilise the independent column density measures from the HI4PI survey \citep{HI4PI2016} and the XMM-Newton X-ray column density values from \citet{Rosine2016}.

\subsubsection{HI4PI} \label{2.2.1}

The HI 4-PI Survey (HI4PI) is a full-sky survey of the 1420 MHz line of neutral atomic hydrogen. It combines the Effelsberg-Bonn HI Survey \citep{Kerp2011}, obtained using a 100-m radio telescope, and the Galactic All-Sky Survey \citep{McClure-Griffiths2009}, carried out using a 64 m Parkes dish. The HI4PI survey has an angular resolution (FWHM) of 16.2' and RMS sensitivity of 43 mK. Also, improving upon previously existing HI datasets, this survey has full spatial sampling. For our work, we use the dataset for the atomic neutral hydrogen (HI) column density map derived from HI4PI. The dataset provides a $V_{LSR}$ range from $\rm \sim -600$ to $+600 \; km \; s^{-1}$, which is more than adequate to provide coverage of the NPS. More description of the survey can be found in \citet{HI4PI2016}.

\subsubsection{XMM Newton} \label{2.2.2}

The X-ray Multi-Mirror Mission/High Throughput X-ray Spectroscopy Mission detects X-ray emission using three telescopes equipped with five imaging cameras and spectrometers which operate simultaneously. Further details of the mission can be found in \citet{XMM2012}.
In this work, we use the column density values for fields towards the NPS as reported in \citet{Rosine2016}, wherein pointed observations towards the southern terminus of the NPS were made with XMM-Newton. The \citet{Rosine2016} work follows the procedure described in \citet{Snowden2008} and calibration from \citet{kuntz2008} to process the XMM-Newton data. They fit the spectra using models that include three thermal emission components, wherein the NPS is represented by an absorbed hot component. They fit for the NPS fluxes and associated absorption column densities, amongst other parameters. Further details regarding the data and the spectral analysis method used can be found in \citet{Rosine2016}.

\section{METHODS USED}

 Our methodology for determining the distance to the NPS builds upon the procedure described in \citet{Zucker2019}, which in turn builds on the methodology used in \citet{Schlafly2014}. We utilise near-infrared (NIR) and optical photometry to infer the distances and extinction to stars towards the NPS, incorporating Gaia DR2 parallax measurements \citep{Lindegren2018} when available. In this section, we summarise the procedure we use to obtain the extinction and distance to each star and fit for the distribution of cumulative extinction as a function of distance (extinction profiles) for the fields towards the NPS, along with any modification made to the model used in \citet{Zucker2019}. \smallskip
 
 \subsection{Fitting for extinction and distance to each star}
 
Following \citet{Green2014, Green2015, Green2018}, we model the observed apparent magnitudes of each star as a function of its extinction, distance, intrinsic stellar type, and $R_V$:

\begin{equation}
\textbf{\textit{m}}_{\rm mod} = \boldsymbol{M}_{\rm int}(M_r, {\rm [Fe/H]}) + A_V \times (\boldsymbol{R} + R_V \times \boldsymbol{R}') + \mu
\end{equation}
where $\mathbf{M}_{\rm int}$ is the set of intrinsic absolute magnitudes for the star as a function of stellar type, $A_V$ is the dust extinction in visual magnitudes, $R_{V}$ ($=A_{V}$/E(B-V)) is the differential extinction, $\boldsymbol{R}$ and $\boldsymbol{R}'$ characterise the overall reddening as a function of magnitude \citep[see][]{Schlafly2016}, and $\mu$ is the distance modulus. The stellar templates we use to model the 2MASS and PS1 photometry are identical to those described in \citet{Green2018}. These empirical models are functions of a vector that tracks the effect of metallicity (Fe/H]) and the absolute r-band magnitude ($M_{r}$) as a function of the star's intrinsic colour in Pan-STARRS1. Further details on these parameters can be found in \citet{Zucker2019}. 

Thus, our model to estimate the per-star distance-extinction includes five parameters: distance modulus $\mu$, overall extinction $A_{V}$, attenuation curve shape $R_{V}$, metallicity [Fe/H] and the PS1 r-band absolute magnitude $M_{r}$. Based on this model, the posterior probability that a set of observed magnitudes \boldsymbol{{$\widehat{m}$}} is consistent with our modelled photometry $\textbf{m}_{\rm mod}(\boldsymbol{\theta})\equiv \textbf{m}_{\rm mod}(M_{r},[Fe/H],A_{V},R_{V},\mu)$ and Gaia DR2 parallax measurements (\textbf{$\hat\varpi$}) is given by:

\begin{equation}
    P(\boldsymbol{\theta}|\boldsymbol{{\widehat{m}}},\hat{\varpi}) \propto \mathcal{L}
(\boldsymbol{{\widehat{m}}}|\boldsymbol{\theta})\ \mathcal{L}
(\boldsymbol{\hat{\varpi}}|\mu)\ \pi(\theta)
\end{equation}
We assume that the likelihood function is independent and is Gaussian in each band. Also, we model the joint prior as: 
\begin{equation}
    \pi(M_{r},[Fe/H],A_{V},R_{V},\mu)=\pi(A_{V})\ \times\ \pi(R_{V})\ \times\ \pi(M_{r})\ \times\ \pi(\mu,[Fe/H])
\end{equation}

We take the prior in $A_{V}$ to be flat between 0 and 12 mag. We take the prior for $R_{V}$ as a normal function with mean of 3.32 and standard deviation of 0.18 \citep[based on][]{Schlafly2016}. The prior on $M_{r}$ is based on PS1 measurements taken from \citet{Green2014}. The joint prior on distance $\mu$ and metallicity [Fe/H] uses the 3D galactic model of \citet{Green2014}. We use the \brutus \footnote{Github link: \url{https://github.com/joshspeagle/brutus}; Zenodo link: \url{https://doi.org/10.5281/zenodo.3348370}} code (Speagle et al. 2019, in prep) to derive the posterior for each source using linear optimisation and brute-force methods. Random samples from these posteriors are then used to calculate marginalised 2D posteriors in $\mu$ and $A_{V}$. 

\subsection{Fitting for line-of-sight extinction profile}

\subsubsection{Southern Terminus} \label{3.2.1}
Our model for how the dust varies as a function of distance is based on \citet{Schlafly2014}, which parameterises the extinction towards a sightline as being caused by a single thin dust screen at the cloud distance modulus $\mu_{c}$. However, in this work for the southern terminus, we find that a single cloud model is unable to account for all the dust structure along the line-of-sight, because of cloud-cloud confusion near the Galactic plane. Hence we modify the model to include multiple dust screens at distances $\mu_{c_i}$'s, where i $\in$ $\mathbb{N}$
 is the dust screen number,
 \begin{equation}\label{eqn:3}
     A_{V}(\mu)=\left\{
  \begin{array}{@{}ll@{}}
    \textit{f} & \text{if } \mu \leq \mu_{c_1} \\
    A_{V_i}(\mu)& \text{if } \mu_{c_i} \leq \mu \leq \mu_{c_{i+1}}          
  \end{array}\right.
\end{equation}
where the foreground extinction is characterised by a constant \textit{f}, the distance to each dust screen is characterised by {$\mu_{c_i}$}, and the extinction at each dust screen is characterised by $A_{v_i}$. We account for variations within a given spatial region in the foreground $(s_{\rm fore})$ and background $(s_{\rm back})$ of each sightline. The parameter $P_{b}$ accounts for the outlier stars in each sightline, by modelling the fraction of stars inconsistent with our model. For further details regarding the parameters, refer to \citet{Zucker2019}. 

Let $ \boldsymbol{\alpha}=\{{\mu_{c_1},...., \mu_{c_n}, A_{v_1},...., A_{v_n}, f, s_{\rm fore}, s_{\rm back}, P_{b}}\} $ be used to parameterise the extinction profile, where $n$ is the number of dust screens/clouds. To obtain the likelihood of our line-of-sight model parameters $\boldsymbol{\alpha}$, we take the product of the line-integral over the extinction profile (defined by Equation \ref{eqn:3}) through the individual stellar posteriors. See Equations 6-14 in \citet{Green2019} or Equations 6-9 in \citet{Zucker2019} for the derivation. 


In this work, we fit for five dust screens, which is arbitrary. For the  lower latitude sightlines, we are more concerned with predicting the range of distances where we can account for most of the dust rather than precise distance estimates for clouds within that range, which is difficult at low latitudes due to the confusion towards the Galactic plane. We use the nested sampling code \dynesty \footnote{\url{https://github.com/joshspeagle/dynesty}} \citep{Speagle2019} to sample for the following free parameters: the foreground extinction f, the foreground smoothing $\rm s_{fore}$, the background smoothing $\rm s_{back}$, the outlier fraction $P_{b}$, the set of cloud distances \{$\mu_{c_1},\mu_{c_2},\mu_{c_3},\mu_{c_4},\mu_{c_5}$\}, and the set of cloud extinctions \{$A_{v_1}$, $A_{v_2}$, $A_{v_3}$, $A_{v_4}$, $A_{v_5}$\}, used in our model. Here $\mu_{c}$'s and $A_{v}$'s are the main free parameters of interest with the other parameters providing the freedom to fit the correct distances and extinctions. The priors that we use for these parameters are the same as those used in \citet{Zucker2019}.
We use the `rwalk' (random walk) sampling strategy of the \dynesty. The detailed sampling \dynesty\  setup used is shown in Appendix \ref{A}. 

\subsubsection{High Latitude portion of NPS}\label{3.2.2}
For the NPS high latitude fields ranging from latitudes of $26^{\circ}$ to $55^{\circ}$, the column density and the number of foreground stars is low. Here, we find that a single dust screen model is able to account for the dust extinction robustly. We adopt a slightly different parameterization of the extinction to the cloud in the single cloud model. Instead of inferring a flat extinction to the cloud $Av_1$, we allow the extinction to vary from star to star based on a spatial template given by \textit{Planck} \citep{Planck2014}. We allow the overall normalisation N with respect to \textit{Planck} to vary to account for the scale difference between our per-star "Bayestar" derived extinction and the \textit{Planck}-derived extinction. Using a spatial template helps to regularise our fit for the high latitude sightlines which have low column densities.   
 The extinction parameter $A_{V}$ for these fields is modelled as:
 \begin{equation}
     A_{v}(\mu)=\left\{
  \begin{array}{@{}ll@{}}
    \textit{f} & \text{if } \ \mu < \mu_{C} \\
    N \times C^{i} & \text{if } \ \mu \geq \mu_{C}   
  \end{array}\right.
\end{equation}
where $\mu_{C}$ is the distance modulus of the dust screen, and $C^{i}$ is the \textit{Planck}-based extinction towards an individual star i. We follow the same likelihood function described for the multi-cloud fits, except we allow the extinction to vary on a star by star basis. This is identical to the parameterization of the extinction used in \citet{Zucker2019} and in \citet{Schlafly2014}.

\subsection{Sample Selection}
 
We use two different selection criteria for stars following \citet{Schlafly2014}: `M-dwarf only' and `all stellar types'. The `M-dwarf only' cut is used to prevent the small number of foreground stars from being overwhelmed by a large number of background stars for nearby dust features, leading to uncertain distances. We select M-dwarf stars based on colour and magnitude cuts along the reddening vector \citep[see Eq. 14 and 15 in][]{Zucker2019}. 
For the sightlines towards the southern terminus, we use stars of all stellar types to have an unbiased sample. We remove stars whose integrated reddening is consistent with being < 0.15 mag \citep[based on \textit{Planck} reddening estimates at 353 GHz;][]{Planck2014} since these regions are unlikely to be informative to the fit. This is the same as the cut used in \citet{Zucker2019}. To limit the number of stars and as another quality cut, we select only those stars which are detected in all the eight Pan-STARRS1 bands. \\\\
For high latitude sightlines, we first checked whether clouds were evident at far distances by keeping both M-dwarf and non M-dwarf type stars. However, we find that there is no significant increase in the integrated extinction beyond $\sim$ 400 pc. Further evidence in support of this assumption can be seen in \S \ref{4.2.2}, where we show that we are able to account for nearly all the integrated column density along the NPS as being nearby. Thus, to maximise the number of foreground stars in our fit and obtain more accurate distances, we consider only the M-dwarf stars for our analysis. Also, we select only those stars which are detected in at least four bands in total \citep[similar to that used in][]{Zucker2019}. We mask out all stars with \textit{Planck}-based \textit{E(B-V)} < 0.03 mag. We had to pick a lower number as compared to the southern terminus because at very high latitudes, the \textit{E(B-V)} and the number of foreground stars is low. Also, at the same time we do not want to include stars that are incapable of informing where a jump in reddening occurs. 

Finally, for all the fields, we also ignore stars with low chi-square fit values:
\begin{equation}
    P(\chi^{2}_{\rm{n_{bands}}} >\chi^{2}_{\rm{best}}) < 0.01
\end{equation} 
where $n_{\rm{bands}}$ is the number of bands of photometry the star is observed in. 

\subsection{Sightlines Used}

\begin{figure}
	\includegraphics[width=99mm]{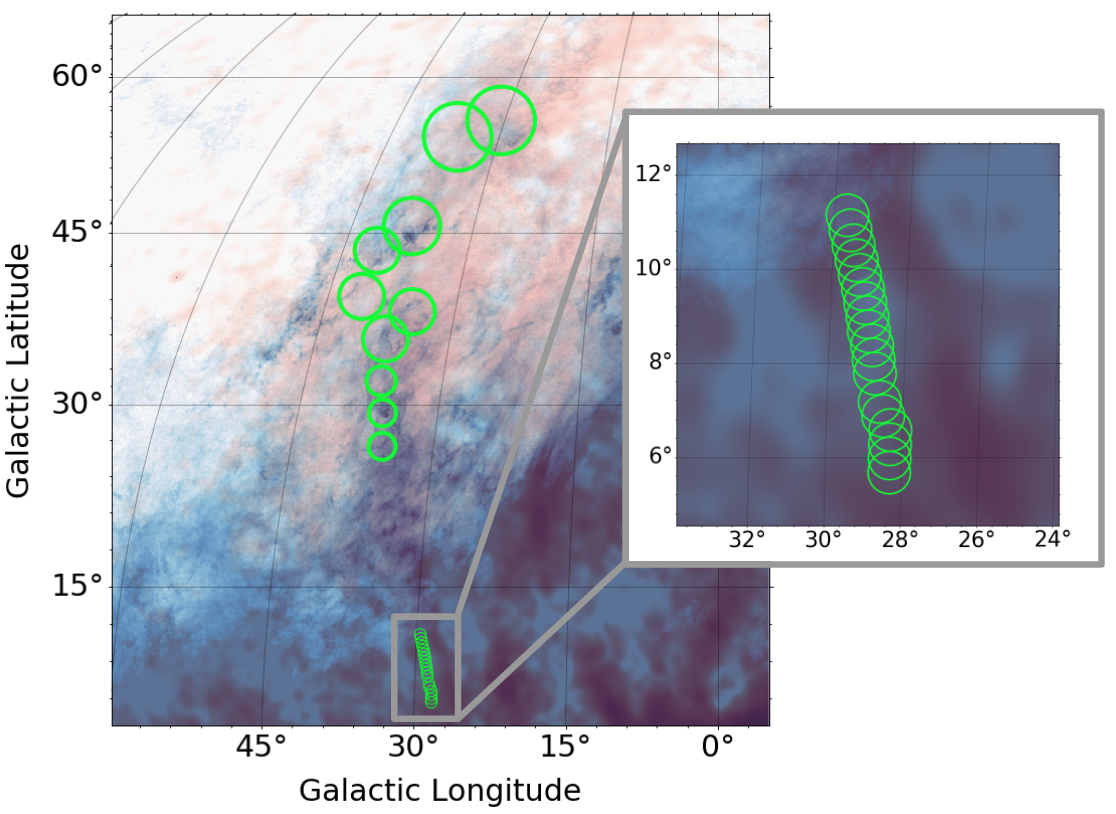}
    \caption{The fields used along the southern terminus of the NPS ($b<11^{\circ}$), and the higher latitudes ($b>26^{\circ}$) are shown in green circles. In the background, the Planck \textit{E(B-V)} map is shown in blue and the Planck 30 GHz polarised intensity map is shown in red}. 
    \label{fig:fields}
\end{figure}

For the first section of the work, we focus on the southern terminus of the NPS. \citet{Rosine2016} provides distance constraints to the X-ray rich southern terminus of the NPS by using column density measurements from \textit{XMM-Newton} data. For the first section of this work, we pick the same 19 fields used in their work (Fig \ref{fig:fields}) towards the southern NPS terminus. The fields have a beam width of $0.5^{\circ}$ and range from Galactic latitudes of $5.6^{\circ}$ to $11.1^{\circ}$.  \\\\

In the second section of the paper, we target higher latitude sightlines towards the NPS, ranging from latitudes of $26.4^{\circ}$ to $55.5^{\circ}$. Figure \ref{fig:fields} shows the ten fields we targeted, superimposed on a combined map of the Planck \textit{E(B-V)} reddening (blue) and the Planck 30 GHz polarised intensity (red). Fields were preferentially chosen towards the western edge of the NPS as seen in polarised intensity from Planck at 30 GHz (see Figure \ref{fig:fields}), as this edge contains appreciable dust emission ($E(B-V) > 0.05$ mag), compared to the eastern edge. This is also the portion of the NPS which is more strongly visible in HI, in comparison to the most prominent section of the NPS in radio continuum, which contains ionised gas \citep[see e.g. discussion in][]{Heiles1980}. The coordinates and beam radius of the ten fields used are provided in Table \ref{table high}. Due to the need to maintain enough foreground stars as we target higher and higher latitudes (toward sightlines with lower and lower stellar densities), we incrementally increase the beam radius from $1.2^\circ$ at a latitude of $26^\circ$ to $3^\circ$ at a latitude of $54^\circ$. 



\begin{landscape}

\begin{table} 
\begin{center}
\caption{A summary of results for the southern terminus NPS sightlines. Columns (1) and (2) give the longitude and latitude of the centre of each field in degrees. Columns (3)-(7) show the distances obtained for the five dust-screens for each sightline.  The errors represents the statistical uncertainties. Columns (8), (12), (14), (16), (18) give the percentage of column density accounted for in each jump as compared to that of the total column density predicted from XMM-Newton data, taken from \citet{Rosine2016}. Columns (9), (11), (13), (15), (17) give the percentages of the total integrated extinction compared to that calculated using the Planck reddening map derived from the $\tau$353GHz dust optical depth map \citep{Planck2014}. A machine readable version of this table is available on the \href{https://dataverse.harvard.edu/dataverse/NPS}{\Dataverse} (doi:10.7910/DVN/UDYNZJ).} 


\begin{tabular}{ccccccccccccccccc}

\hline
l & b &  $D_1$ & $D_2$ & $D_3$  & $D_4$ & $D_5$  & \multicolumn{2}{c}{\% $N_{H}$-1} & \multicolumn{2}{c}{\% $N_{H}$-2} &  \multicolumn{2}{c}{\% $N_{H}$-3} & \multicolumn{2}{c}{\% $N_{H}$-4} & \multicolumn{2}{c}{\% $N_{H}$-5} \\

&&&&&&& XMM& \textit{Planck}& XMM& \textit{Planck}& XMM& \textit{Planck}& XMM& \textit{Planck}& XMM& \textit{Planck} \\
$\circ$&$\circ$& pc & pc& pc &pc& pc && && &&&&&& \\\\
$(1)$&(2)&(3) &(4)& (5) &(6)& (7) &(8)&(9)&(10)&(11)&(12)&(13)&(14)&(15)&(16)&(17)  \\\\
\hline
28.4 & 5.6 & $380^{+31}_{-48}$ & $439^{+17}_{-9}$ & $587^{+32}_{-22}$ & $2427^{+234}_{-196}$ & $3753^{+118}_{-110}$ & $47^{+6}_{-5}$\% &$35^{+4}_{-3}$\%&  $105^{+4}_{-5}$\% &$79^{+3}_{-3}$\% &  $135^{+1}_{-1}$\%&$102^{+1}_{-1}$\%&  $144^{+2}_{-2}$\%&$109^{+1}_{-1}$\%&  $156^{+1}_{-1}$\% &$118^{+0}_{-0}$\\\\
28.4 & 5.9 & $348^{+80}_{-47}$ & $430^{+38}_{-54}$ & $520^{+231}_{-87}$ & $977^{+705}_{-216}$ & $2331^{+294}_{-332}$ & $92^{+55}_{-25}$\% &$50^{+30}_{-14}$\%& $170^{+17}_{-85}$\% &$94^{+9}_{-47}$\%&  $198^{+9}_{-15}$\% &$109^{+5}_{-8}$\%&  $210^{+5}_{-4}$\% &$116^{+3}_{-2}$\%&  $226^{+2}_{-2}$\% &$124^{+1}_{-1}$\%\\\\
28.4 & 6.2 & $191^{+18}_{-57}$ & $290^{+119}_{-18}$ & $434^{+168}_{-11}$ & $681^{+25}_{-23}$ & $1608^{+67}_{-69}$ & $38^{+35}_{-11}$\% &$22^{+21}_{-6}$\%&  $85^{+70}_{-8}$\% &$50^{+41}_{-5}$\%&  $170^{+7}_{-6}$\% &$100^{+4}_{-3}$\%& $219^{+1}_{-2}$\% &$129^{+1}_{-1}$\%& $236^{+1}_{-1}$\% &$139^{+0}_{-0}$\%\\\\
28.4 & 6.5 & $242^{+10}_{-164}$ & $308^{+64}_{-53}$ & $471^{+6}_{-9}$ & $788^{+67}_{-75}$ & $2008^{+117}_{-90}$ &$70^{+14}_{-46}$\% &$52^{+11}_{-34}$\%& $96^{+4}_{-4}$\% &$71^{+3}_{-3}$\%& $147^{+4}_{-5}$\% &$109^{+3}_{-4}$\%&  $172^{+7}_{-6}$ \% &$127^{+1}_{-1}$\%&  $188^{+1}_{-1}$\% &$139^{+0}_{-0}$\%\\\\
28.6 & 6.8 & $140^{+47}_{-52}$ & $266^{+35}_{-17}$ & $474^{+25}_{-50}$ & $748^{+50}_{-33}$ & $2655^{+180}_{-250}$ & $37^{+10}_{-10}$\% &$37^{+10}_{-10}$\%&  $76^{+6}_{-7}$\% &$76^{+6}_{-7}$\%&  $109^{+4}_{-4}$\% &$108^{+4}_{-4}$\%&  $127^{+1}_{-1}$\% &$127^{+1}_{-1}$\%&  $137^{+1}_{-1}$\% &$137^{+1}_{-1}$\%\\\\
28.7 & 7.1 &  $258^{+6}_{-10}$ & $368^{+15}_{-16}$ & $557^{+12}_{-15}$ & $1051^{+52}_{-57}$ & $3842^{+102}_{-200}$ & $62^{+4}_{-4}$\% &$58^{+4}_{-4}$\%&  $91^{+3}_{-3}$\% &$85^{+3}_{-3}$\%&  $119^{+2}_{-2}$\% &$111^{+1}_{-2}$\%&  $128^{+1}_{-1}$\% &$119^{+1}_{-1}$\%&  $141^{+0}_{-0}$\% &$131^{+0}_{-0}$\%\\\\
28.8 & 7.4 & $269^{+14}_{-17}$  & $422^{+31}_{-29}$ & $575^{+21}_{-17}$ & $979^{+55}_{-69}$ & $3351^{+160}_{-171}$ & $107^{+5}_{-7}$\% &$53^{+3}_{-4}$\% &  $154^{+7}_{-8}$\% &$76^{+4}_{-4}$\%&  $203^{+3}_{-3}$\% &$101^{+1}_{-1}$\%& $219^{+1}_{-1}$\% &$109^{+1}_{-1}$\%&  $243^{+1}_{-1}$\% &$121^{+1}_{-1}$\%\\\\
28.8 & 7.7 & $158^{+124}_{-39}$ & $284^{+197}_{-31}$ & $568^{+38}_{-23}$ &  $898^{+25}_{-30}$ & $3121^{+470}_{-214}$ &$28^{+26}_{-12}$\%  &$28^{+31}_{-48}$\%& $59^{+16}_{-6}$\% &$60^{+31}_{-48}$\% & $91^{+2}_{-2}$\% & $93^{+31}_{-48}$\% & $103^{+2}_{-2}$\% &$105^{+31}_{-48}$\% & $113^{+0}_{-0}$\% & $115^{+31}_{-48}$\%\\\\
28.9 & 8.0 & $142^{+71}_{-54}$ & $351^{+20}_{-18}$ & $594^{+22}_{-38}$ & $865^{+30}_{-43}$ & $2913^{+230}_{-205}$ & $44^{+6}_{-6}$\% &$36^{+5}_{-5}$\%&  $74^{+3}_{-3}$\% &$61^{+3}_{-2}$\%&  $114^{+1}_{-4}$\% &$94^{+3}_{-3}$\%&  $132^{+1}_{-1}$\% &$108^{+1}_{-1}$\%&  $144^{+1}_{-1}$\% &$118^{+0}_{-0}$\%\\\\
29.0 & 8.3 & $299^{+21}_{-15}$ & $505^{+27}_{-50}$ & $762^{+76}_{-232}$ & $2113^{+592}_{-1319}$ & $2976^{+539}_{-228}$ & $87^{+6}_{-6}$\% &$62^{+4}_{-5}$\%&  $125^{+5}_{-25}$\% &$90^{+4}_{-18}$\%&  $150^{+2}_{-23}$\% &$108^{+2}_{-17}$\%&  $156^{+8}_{-4}$\% &$112^{+5}_{-3}$\%&  $167^{+1}_{-1}$\% &$120^{+1}_{-1}$\%\\\\
29.0 & 8.6 & $275^{+23}_{-23}$ & $344^{+136}_{-15}$ & $531^{+257}_{-25}$ & $920^{+80}_{-58}$ & $2414^{+303}_{-151}$ &  $58^{+17}_{-8}$\% &$53^{+16}_{-7}$\%& $81^{+18}_{-3}$\% &$75^{+16}_{-3}$\%&  $102^{+6}_{-2}$\% &$93^{+6}_{-2}$\%&  $118^{+1}_{-1}$\% &$108^{+1}_{-1}$\%& $129^{+1}_{-0}$\% &$118^{+1}_{-1}$\%\\\\
29.1 & 8.9 & $308^{+8}_{-9}$ & $338^{+37}_{-17}$ & $613^{+71}_{-149}$ & $907^{+82}_{-20}$ & $3528^{+246}_{-161}$ & $82^{+8}_{-9}$\% &$61^{+6}_{-7}$\%&  $100^{+5}_{-5}$\% &$74^{+4}_{-5}$\%&  $122^{+8}_{-5}$\% &$91^{+6}_{-4}$\%&  $147^{+1}_{-1}$\% &$109^{+1}_{-1}$\%&  $159^{+1}_{-1}$\% &$118^{+1}_{-1}$\%\\\\
29.2 & 9.2 & $294^{+15}_{-8}$ & $327^{+8}_{-19}$ & $790^{+64}_{-314}$ & $1101^{+201}_{-243}$ & $2963^{+185}_{-326}$ &$62^{+6}_{-12}$\% &$55^{+5}_{-11}$\%&  $108^{+3}_{-5}$\% &$96^{+3}_{-4}$\%&  $120^{+8}_{-5}$\% &$107^{+3}_{-9}$\%&  $128^{+2}_{-1}$\% &$113^{+2}_{-1}$\%&  $138^{+1}_{-1}$\% &$122^{+1}_{-1}$\%\\\\
29.2 & 9.5 & $297^{+10}_{-10}$ & $330^{+15}_{-20}$ & $812^{+37}_{-459}$ & $997^{+315}_{-143}$ & $2792^{+452}_{-490}$ & $124^{+16}_{-14}$\% &$61^{+8}_{-7}$\%&  $193^{+4}_{-35}$\% &$95^{+2}_{-17}$\%&  $220^{+8}_{-5}$\% &$109^{+3}_{-11}$\%&  $227^{+3}_{-2}$\% &$112^{+1}_{-1}$\%& $249^{+2}_{-2}$\% &$123^{+1}_{-1}$\%\\\\
29.4 & 9.8 & $96^{+178}_{-22}$ & $286^{+19}_{-10}$ & $493^{+22}_{-51}$ & $964^{+112}_{-153}$ & $2985^{+176}_{-169}$ &$36^{+43}_{-6}$\% &$32^{+39}_{-5}$\%&  $84^{+5}_{-5}$\% &$76^{+4}_{-5}$\%&  $104^{+3}_{-4}$\% &$94^{+2}_{-3}$\%&  $114^{+1}_{-1}$\% &$103^{+1}_{-1}$\%&  $129^{+1}_{-1}$\% &$116^{+1}_{-1}$\%\\\\
29.4 & 10.1 & $211^{+20}_{-24}$ & $284^{+36}_{-41}$ & $581^{+87}_{-260}$ & $1185^{+384}_{-473}$ & $2984^{+163}_{-217}$ &$37^{+43}_{-6}$\% &$39^{+8}_{-9}$\%&  $71^{+6}_{-26}$\% &$74^{+6}_{-28}$\%&  $87^{+3}_{-10}$\% &$91^{+3}_{-11}$\%&  $93^{+2}_{-1}$\% &$97^{+2}_{-1}$\%&  $103^{+1}_{-1}$\% &$108^{+1}_{-1}$\%\\\\
29.6 & 10.4 & $108^{+43}_{-36}$ & $288^{+23}_{-27}$ & $552^{+34}_{-49}$ & $1786^{+67}_{-67}$ & $3917^{+49}_{-117}$ & $36^{+5}_{-5}$\% &$43^{+6}_{-6}$\%&  $58^{+3}_{-4}$\% &$68^{+4}_{-4}$\%&  $75^{+1}_{-1}$\% &$89^{+1}_{-1}$\%&  $82^{+1}_{-1}$\% &$97^{+1}_{-1}$\%&  $89^{+1}_{-1}$\% &$106^{+1}_{-1}$\%\\\\
29.6 & 10.8 & $251^{+21}_{-150}$ & $412^{+35}_{-158}$ & $523^{+165}_{-96}$ & $1739^{+219}_{-867}$ & $3581^{+304}_{-304}$ &$60^{+6}_{-27}$\% & $66^{+6}_{-29}$\%&  $75^{+5}_{-14}$\% &$81^{+5}_{-15}$\%&  $82^{+1}_{-1}$\% &$89^{+1}_{-1}$\% &  $87^{+2}_{-2}$\% &$94^{+2}_{-3}$\% &  $97^{+1}_{-1}$\% &$105^{+1}_{-1}$\%\\\\
29.8 & 11.1 & $99^{+73}_{-23}$ & $348^{+155}_{-215}$ & $662^{+127}_{-209}$ & $1012^{+476}_{-257}$ & $3301^{+215}_{-296}$ &  $54^{+8}_{-11}$\% &$55^{+8}_{-11}$\%&  $66^{+6}_{-5}$\% &$67^{+6}_{-5}$\%&  $78^{+5}_{-8}$\% &$79^{+5}_{-8}$\% &  $86^{+2}_{-1}$\% &$87^{+2}_{-1}$\%& $97^{+2}_{-2}$\% &$99^{+2}_{-2}$\%\\\\
\hline
\end{tabular} \label{Table1}
\end{center}
\end{table}

\end{landscape}

\section{Results}
\subsection{Distances to the Southern Terminus} \label{4.1}

\subsubsection{Extinction Profiles}

\begin{figure*}
    
    \subfloat[][]{\includegraphics[scale = 0.55]{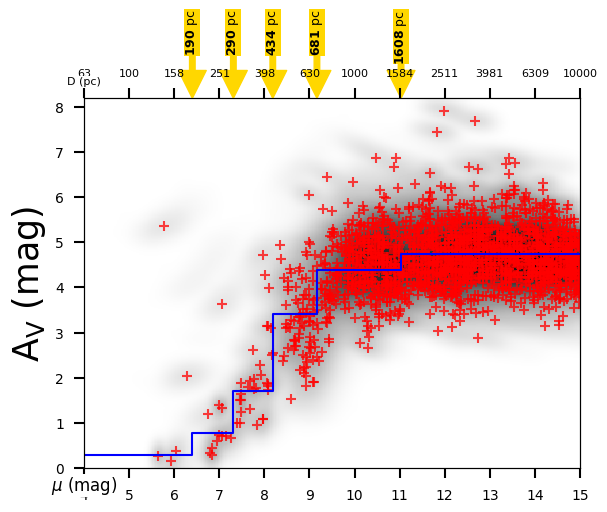}}
    \subfloat[][]{\includegraphics[scale = 0.55]{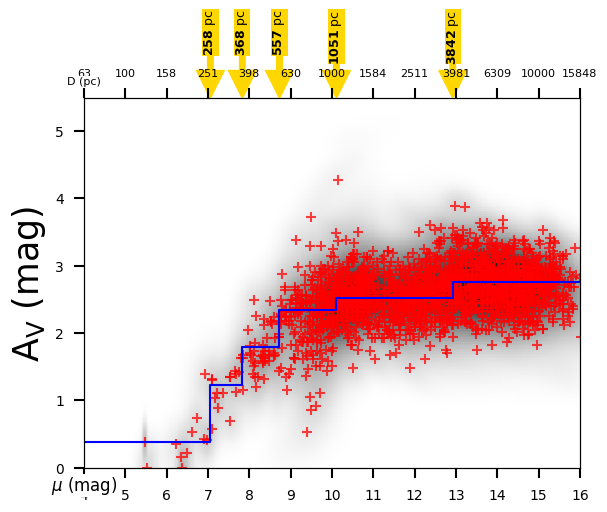}}\\
    \caption{Left: line-of-sight plot for (l,b)=($28.4^{\circ}, 6.2^{\circ})$, Right: (l,b)=$(28.7^{\circ}, 7.1^{\circ})$. The extinction ($A_{V}$) vs distance modulus ($\mu$) plots show the distances to each jump, which are marked with the yellow arrows. The red crosses correspond to the most likely extinction and distance for each star. We note that the dust jumps do not necessarily correspond to discrete clouds. At low latitudes, we are more concerned with predicting the range of distances where we can account for most of the dust associated with the NPS,  rather than precise distance estimates for clouds within that range, which is difficult to constrain due to cloud confusion in the Galactic plane.}   
    \label{fig:south}
\end{figure*}

We obtain the line-of-sight extinction profiles towards each field along the southern terminus using the procedure described in Section \ref{3.2.1}. The line-of-sight plots for two fields are shown in Figure \ref{fig:south}. Similar plots for all the sightlines mentioned in Table \ref{Table1} can be found on the \href{https://dataverse.harvard.edu/dataverse/NPS}{\Dataverse} (doi:10.7910/DVN/FGQAPG). These extinction ($A_{V}$) vs. distance modulus ($\mu$) plots show the distances to each jump, marked by the yellow arrows. The red crosses correspond to the most likely extinction and distance for each star. The distances at which the dust screens fitted for are located are provided in Table \ref{Table1}. We find that the jumps/dust screens are located within a range of 100 pc to 3 kpc for all the sightlines. However, as will be shown in the next subsection, negligible column density is accounted for by the dust screens located beyond 1 kpc. The trace plots and corner plots obtained for for all the sightlines mentioned in Table \ref{Table1}, showing the reliability of the model used and the results obtained is available on the \href{https://dataverse.harvard.edu/dataverse/NPS}{\Dataverse} (doi:10.7910/DVN/FGQAPG).   

\subsubsection{Comparison with XMM-Newton column density and Planck reddening maps}\label{4.1.2}

For each of our dust screens shown in Table \ref{Table1}, we determine the fraction of the total column density associated with the NPS \citep[derived from XMM-Newton data;][]{Rosine2016} that can be accounted for at each distance.  The XMM-derived column densities
for the NPS should represent the total integrated column density
foreground to the structure at infinity, so the distance at which we
reach 100 per cent should represent the near-side distance of the
structure.

Direct comparison of our inferred model parameters (the amount of extinction at different distances) and the XMM-Newton column densities from \citet{Rosine2016} requires the adoption of different conversion coefficients. For the values in Table \ref{Table1}, we convert the inferred extinction to a reddening using the mean $R_{V}$ of stars in the field (determined by our modelling described in \S \ref{3.2.1}). Then, to convert the reddening values to column density, we use the conversion factor of $5.8 \times 10^{21} \rm \; cm^{-2} \; mag$ reported in \citet{Bohlin1978}. We also repeat our calculations using a conversion factor of $4.0 \times 10^{21} \rm \; cm^{-2} \; mag$ reported in \citet{Rosine2016} and $8.8 \times 10^{21} \rm \; cm^{-2} \; mag$ reported in \citet{Lenz2017}. We find that our results are insensitive to the adoption of the conversion factor used. 

We also calculate the extinction percentage accounted for at each step with respect to the total reddening calculated using the colour excess map from the Planck reddening map \citep{Planck2014}. We report these values in Table \ref{Table1}. To convert to extinction values we adopt $R_{V}$ = 3.1 as mentioned in \citet{Schlafly2014} and \citet{Green2019}.   We find that for a majority of the sightlines (14
of 19 fields), we can account for 100 per cent of the column density
to the NPS from Planck at distances between 400 and 700 pc. Since
any dust associated with the NPS should be captured by Planck, this
suggests that the NPS must lie within 700 pc. However, we cannot
entirely preclude farther distances for some sightlines (e.g. l, b =
$29.8^{\circ}, 11.1^{\circ}$).


\begin{figure*} 
    \subfloat[][]{\includegraphics[scale = 0.4]{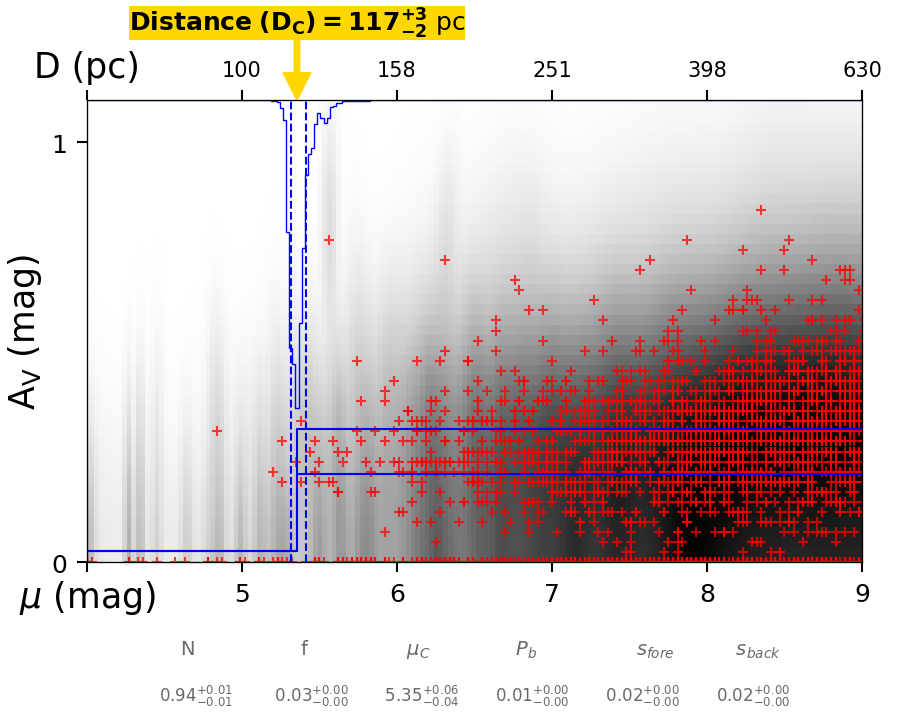}}
    \subfloat[][]{\includegraphics[scale = 0.4]{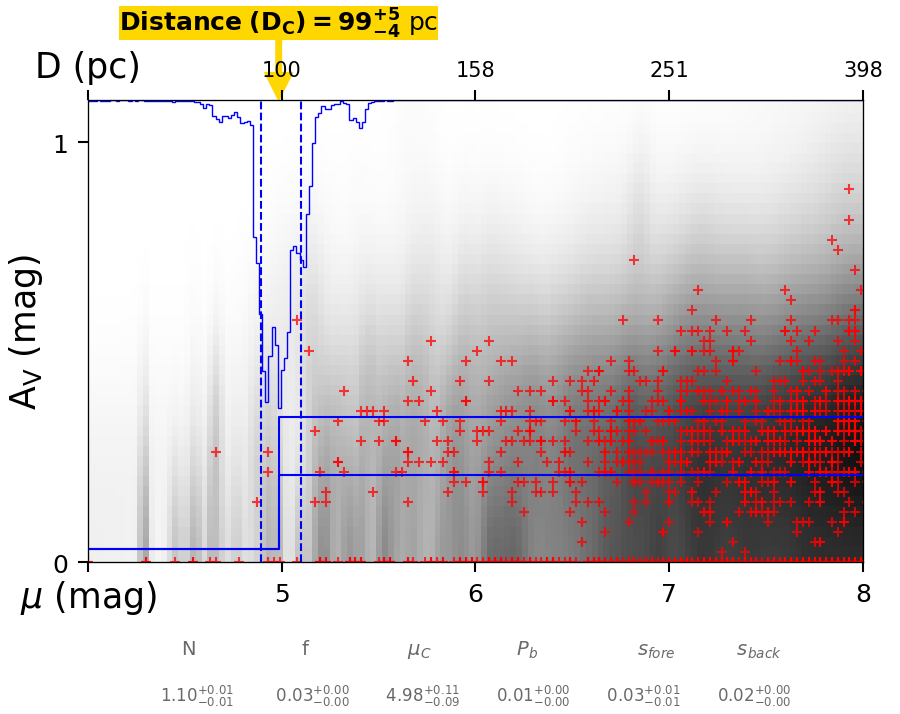}}

    \caption{Left: line-of-sight plot for (l,b)=($37.31^{\circ}, 35.53^{\circ})$, Right: (l,b)=$(40.98^{\circ}, 43.28^{\circ})$, using a single cloud fit for fields towards high latitudes of the NPS. These extinction ($A_{V}$) vs. distance modulus ($\mu$) plots contain the distance to the dust cloud marked with the yellow arrow. The inverted blue histogram at the top shows the probable range of distances to the cloud. The dashed vertical lines correspond to the $16^{th}$ and $84^{th}$ percentile values of the cloud distance. The red crosses correspond to the most likely extinction and distance for each star.}   
    \label{fig:high}
\end{figure*}

\begin{table}
\begin{center}
\caption{A summary of results for the high latitude NPS sightlines. Columns (1) and (2) give the longitude and latitude of each field in degrees. Column (3) provides the radius of the region targeted. Column (4) shows the distance obtained for each sightline. The first set of errors represents the statistical uncertainties, while the second set is the systematic uncertainty , estimated to be 5\% in distance; see \citet{Zucker2019}. Columns (5)-(6) shows the percentage of the total column density we are able to account for by comparison with an independent extinction measure from the HI4PI (\citep{HI4PI2016}). In column (5), we use the conversion factor of $\rm 5.8\times10^{-21}\ cm^{-2}$ \citep{Bohlin1978} while in column (6), we use a factor of $\rm 8.8\times10^{-21}\ cm^{-2}$ \citep{Lenz2017}. A machine readable version of this table is available on the \href{https://dataverse.harvard.edu/dataverse/NPS}{\Dataverse} (doi:10.7910/DVN/I3S7AT).
}
\begin{tabular}{cccccc} 

\hline
l & b & Radius & Distance & \%$\rm {NH_{Bohlin}}$ & \%$\rm {NH_{Lenz}}$  \\\\
$\circ$ & $\circ$ & $\circ$ & pc && \\\\ 
(1) & (2)& (3) & (4) & (5) & (6)\\

\hline
35.5 & 26.4 & 1.2 & $135^{+2}_{-22}\pm7$  &$92^{+15}_{-18}$\% & $144^{+24}_{-27}$ \% \\\\
35.9 & 29.2 & 1.2 & $137^{+2}_{-2}\pm7$  &$94^{+29}_{-29}$\% & $148^{+46}_{-46}$ \% \\\\
36.8 & 31.9 & 1.3 &$131^{+1}_{-1}\pm7$ &$99^{+19}_{-18}$\% & $156^{+29}_{-28}$ \% \\\\
37.3 & 35.5 & 2.0 & $117^{+3}_{-2}\pm6$ &$87^{+20}_{-16}$\% & $137^{+32}_{-25}$ \% \\\\
35.0 & 37.9 & 2.0 &$126^{+5}_{-4}\pm6$& $98^{+18}_{-16}$\% & $153^{+28}_{-26}$ \% \\\\
41.3 & 39.9 & 2.0 &$86^{+5}_{-7}\pm4$& $83^{+18}_{-23}$\% & $138^{+36}_{-28}$ \% \\\\
41.0 & 43.3 & 2.0 &$99^{+4}_{-4}\pm5$& $103^{+29}_{-23}$\% & $163^{+45}_{-36}$ \% \\\\
37.7 & 45.5 & 2.5 &$75^{+3}_{-4}\pm4$& $102^{+37}_{-23}$\% & $160^{+58}_{-37}$ \% \\\\
35.8 & 53.9 & 3.0 & $76^{+9}_{-2}\pm4$ &$94^{+17}_{-14}$\% & $148^{+27}_{-22}$\% \\\\
30.6 & 55.5 & 3.0 & $70^{+5}_{-4}\pm4$ &$77^{+25}_{-19}$\% & $121^{+40}_{-29}$\% \\
\hline

\end{tabular} \label{table high}
\end{center}
\end{table}

\subsection{Distances towards the higher latitudes} \label{4.2}
\subsubsection{Extinction Profiles}

As described earlier, we chose ten fields along the NPS varying from latitudes $26^{\circ}$ to $55^{\circ}$, with the beam radius for each field provided in Table \ref{table high}. We obtain the extinction profiles towards these fields, following the procedure described in \S \ref{3.2.2}. The line-of-sight plots for two fields are shown in Figure \ref{fig:high}. Similar plots for all the sightlines mentioned in Table \ref{table high} can be found on the \href{https://dataverse.harvard.edu/dataverse/NPS}{\Dataverse} (doi:10.7910/DVN/PWKEZ2). These extinction ($A_{V}$) vs. distance modulus ($\mu$) plots mark the distance to the dust cloud with the yellow arrow. The dashed vertical lines correspond to the $\rm 16^{th}$ and $\rm 84^{th}$ percentile values of the cloud distance. The red crosses correspond to the most likely extinction and distance for each star. We find that jumps occur within 140 pc for all the ten fields (see Table \ref{table high}). We find evidence for a distance gradient as a function of latitude, with lower latitude sightlines lying around 130 pc, and higher latitudes sightlines lying around 75 pc. The trace plots and corner plots obtained for for all the sightlines mentioned in Table \ref{table high}, showing the reliability of the model used and the results obtained is available on the \href{https://dataverse.harvard.edu/dataverse/NPS}{\Dataverse} (doi:10.7910/DVN/PWKEZ2).

\subsubsection{Comparison with HI column density}\label{4.2.2}
We obtain an independent column density measure of the NPS at higher latitudes using HI derived column densities over velocities consistent with the NPS. Unsurprisingly, we observe that the NPS is the only feature in the spectrum. We then obtain the spectrum over the area coincident with our sightlines on the plane of the sky using \glue\footnote{\url{glueviz.org}} \citep{glue2015}. We obtain the zeroth-moment HI map for the NPS in different regions by integrating the HI spectral cube obtained from the H4PI Survey (described in \S \ref{2.2.1}) over the velocity range from $ -16.1\ \rm km \; s^{-1}$ to $+23.9\ \rm km \; s^{-1}$, which fully encompasses the NPS spectral feature. That zeroth order moment map is then converted to a HI column density ($N_{HI}$) map using the following conversion factor from \citet{HI4PI2016}: 
\begin{equation}
    N_{HI} \big(\rm cm^{-2}\big) = 1.823 \times 10^{18} \int d\nu \ T_{B}(\nu)\ \big({\rm K\ km\ s^{-1}}\big)
\end{equation} 

To compare the inferred amount of extinction we obtain at the cloud distance to the derived HI column densities, we adopt the same procedure as in \S \ref{4.1.2}.
We obtain the average $N_{HI}$ value by averaging over all the $N_{HI}$ values for the HI4PI pixels for each field of interest. We convert the extinction associated with the cloud (based on our fits) to HI column density using the conversion factor reported in \citet{Bohlin1978}. We then obtain the percentage of column density we can account for using our single-step extinction profiles towards the higher latitudes (see Table \ref{table high}). The results obtained using a different $A_V$ to $N_{HI}$ factor \citep[from][]{Lenz2017} are also reported in Table \ref{table high}. We find that our results are insensitive to the conversion factor used. In Table \ref{table high}, we also report the upper and lower bound to the percentages, calculated from the 16th percentile and 84th percentile respectively, of the extinction values we obtain from our fit. We find that we are able to account for nearly 100\% of the extinction within $\sim$140 pc.  To calculate the contribution of CO column density towards these fields, we use the Planck full-sky CO map (Type-2) \citep{CO2014}. However, we find that there is a negligible difference in the percentage values regardless of whether we take CO into account or not. Thus, we do not consider CO for our analysis of the high-latitude sightlines.  

\section{Discussion}

\begin{figure}
	\includegraphics[width=86mm]{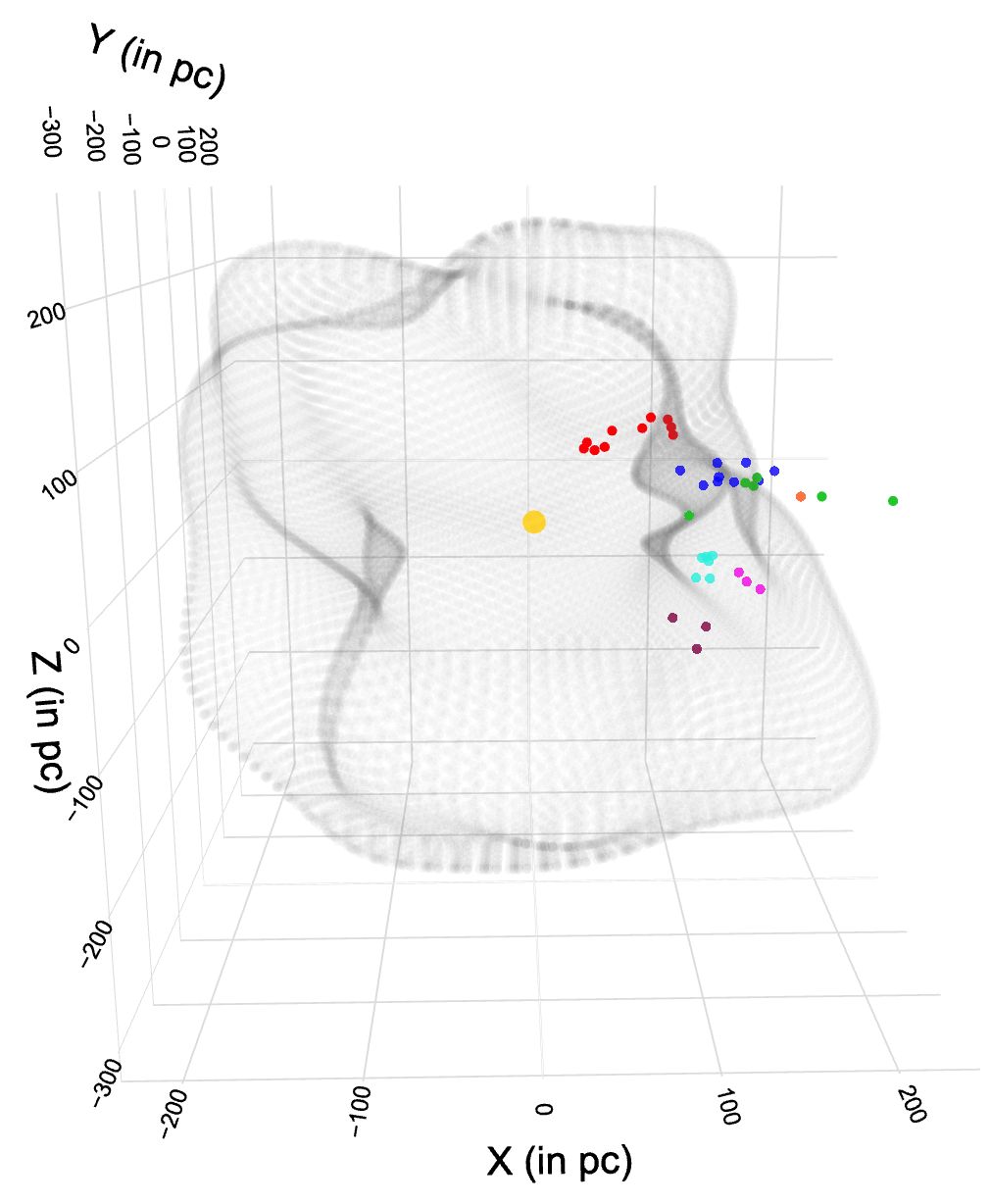}
    \caption{A 3D Cartesian view of the North Polar Spur in context of the Local Bubble and nearby star-forming regions in the Sco-Cen association. The figure shows the NPS sightlines (in red) from Table \ref{table high}. The Sun (in yellow) is at the centre. We have over-plotted the Local Bubble boundary (in grey) \citep[obtained from][]{Pelgrims2020}, and dust clouds associated with the Sco-Cen association: Ophiuchus (in blue), Chamaeleon (in brown), Coalsack (in cyan), Corona Australis (in magenta), Lupus (in green), Barnard 59 (in orange). The distances to the dust clouds are obtained from \citet{Zucker2020}. An interactive version of this figure is available at \href{https://faun.rc.fas.harvard.edu/czucker/Paper_Figures/NPS_3D.html}{https://faun.rc.fas.harvard.edu/czucker/Paper\_Figures/NPS\_3D.html} or in the online version of the published article}. 
    \label{fig:3d}
\end{figure}

The North Polar Spur was originally identified as an intense source of non-thermal radio continuum, and later associated with Loop I, and found to be a steady source of X-ray emission. The origin and distance of the North Polar Spur remains under debate.

One argument is that the NPS is not a local structure, but lies closer to the Galactic centre, with potential association with the Fermi Bubbles \citep{{Sofue2000}, {Sofue2015}, {Sun2014}, {Yofue2019}}. 
One common argument against the association of the NPS with the Fermi bubbles is its lack of a counterpart in the southern hemisphere. However, using numerical simulations, \citet{Sarkar2019} argues that the NPS, Fermi Bubbles and the Loop-I could have a common origin despite this asymmetry, finding that small density variations in the density of the circumgalactic medium can induce this effect in star formation driven wind scenarios towards the Galactic centre.  

Our results do not agree with these claims. Our southern terminus results ($5^\circ<b<11^\circ$) indicate that we can account for $\approx 100\%$ of the column density to the NPS within 700 pc (see \S \ref{4.1}), based on independent measures of the column density of the NPS. For the sightlines in the latitude range $26^{\circ}<b<55^{\circ}$, we are able to account for all the column density to the NPS within 140 pc (see Section \ref{4.2.2}). The slight discontinuity in the distances to the NPS as we go from the southern terminus to higher latitudes can be attributed to the fact that the NPS is one of many structures along the line of sight so close to the Galactic plane. We infer from our distances that the NPS is not associated with the Fermi Bubbles and is not located near the Galactic centre. This is consistent with claims summarised in \citet{Dickinson2018}, which reports that the microwave polarisation data and radio maps indicate little correlation between the Fermi Bubbles and the NPS, as they show almost no interaction and the southern portion of the Fermi Bubbles extends way beyond the NPS. Similarly, \citet{Planck2015} argues that the NPS is not associated with the Galactic Centre based on polarisation maps and geometric constraints.

In Fig \ref{fig:3d} (interactive) we present the NPS sightlines mentioned in Table \ref{table high}, in a Cartesian XYZ coordinate system, where X increases towards the Galactic centre at Galactic longitude $l = 0^{\circ}$, Y increases along the direction of rotation of the Galaxy at Galactic longitude $l = 90^{\circ}$, and Z increases upwards out of the Galactic plane towards the North Galactic pole.  The Sun (in yellow) is at the centre. In the figure, the Local Bubble boundary \citep[obtained from][]{Pelgrims2020} corresponds to the distance of the `inner edge' of the Local Bubble as seen from the Sun and derived from the \citet{Lallement2019} 3D
Galactic interstellar dust map. We refer to \citet{Pelgrims2020} for full details on how the `inner edge' is determined. We also plot dust clouds associated with the Sco-Cen association: Ophiuchus, Chamaeleon, Coalsack, Corona Australis, Lupus, and Barnard 59. The distances to the dust clouds are obtained from \citet{Zucker2020}.  Our Gaia-constrained distances obtained via 3D dust mapping help us narrow down the possible theories for the origin and 3D position of the NPS. The distances towards the higher latitudes give strong evidence behind its possible link with the Scorpius-Centaurus (Sco-Cen) Association. Distances to the various molecular clouds and young stellar objects (YSOs) in the Sco-Cen star-forming regions have been accurately determined by \citet{Zucker2019, Zucker2020} and \citet{Dzib2018}: Ophiuchus (120-140 pc), Lupus I, II, III, IV (156-163 pc), Corona Australis (154 pc), B59 (163 pc), and Chamaeleon I, II (192-198 pc). These distances strongly suggest the association of the NPS, especially at latitudes $\geq 26^{\circ}$, with the Sco-Cen Association. We would also like to emphasise that while some NPS sightlines are consistent with the Local Bubble boundary, most sightlines fall inside the 'inner edge' of the Local Bubble defined in \citet{Pelgrims2020}. There have been other works that suggest the local origin of the NPS. Research on polarised stars with known distances show that the NPS is a local structure within 200 pc \citep{Berdyugin2014,Santos2011}. Also, Faraday tomography of the radio continuum adjacent to the spur region indicates that radio emission from the spur is not Faraday-depolarised; thus it is located within a few hundred parsecs \citep{Sun2014}. \citet{deGues1992} predicted that the NPS is centred at the Sco-Cen OB association, showing that the X-ray remnant toward the NPS could have resulted from star formation triggered by the impact of a shock wave on the Aquila Rift dark cloud. \citet{Wolleben2007} presents a model consisting of
two synchrotron emitting shells, wherein the polarised emission of the NPS is
reproduced by one of these shells. Specifically, \citet{Wolleben2007} proposes that the X-ray emission seen towards the NPS is produced by interaction of the two
shells and that two OB-associations coincide with the centres of the shells. \citet{Dwarkadas2018} argues that the most likely source of the NPS is the Lower Centaurus Crux (LCC) sub-group of the Sco-Cen association. They state that approximately six supernovae in the LCC are responsible for creating the Local Bubble (which is the hot component of the local cavity). \citet{SmithCox2001}
also proposes the Sco-Cen OB association as the source for the supernovae explosions required for the formation of the Local Bubble. On the other hand,  \citet{Frisch1981} and \citet{Breitschwerdt1996} suggest that the Local Bubble was created by the Sco-Cen association, as a blister of Loop I superbubble. Based on our distance constraints, we thus argue that the NPS, especially at higher latitudes, is associated with the Sco-Cen OB association and that its origin is potentially connected with that of the Local Bubble.

\section{Conclusion}
We determine distances to the North Polar Spur using a combination of near-infrared and optical photometry and Gaia DR2 parallax measurements. Using the broadband photometry and Gaia astrometry, we compute distance and extinction to thousands of stars towards the North Polar Spur. We then fit these measurements with multi- and single-cloud dust models to infer the dust distribution toward the southern terminus of the NPS and higher latitudes. We then compute the fraction of the total integrated extinction we are able to account for at different distances, by comparison with independent extinction measures from the HI4PI and XMM-Newton Survey. We are able to provide accurate distance constraints for the NPS and account for nearly 100\% of the total column density within 140 pc for the high latitudes of $26^{\circ}$ to $55^{\circ}$ and within 700 pc for the southern terminus sightlines ranging from $5^{\circ}$ to $11^{\circ}$. Based on our results, we support the claim that the NPS is not associated with the Fermi bubbles at the Galactic Centre, but rather associated with the Sco-Cen OB Association.

\section*{Acknowledgements}

KKD acknowledges Prof. Pankaj Jain for acting as local supervisor for the duration in which the project was carried out at IIT Kanpur. \\

We would like to thank V. Pelgrims for kindly providing the 3D map of the structure of the Local Bubble. \\

We would like to thank Edward Schlafly and Douglas Finkbeiner for invaluable discussions on the methodology leading up to this work. \\

The computations in this paper utilise resources from the Odyssey cluster, which is supported by the FAS Division of Science, Research Computing Group at Harvard University. \\

The visualisation, exploration, and interpretation of data presented in this work was made possible using the glue visualisation software, supported under NSF grant OAC-1739657. \\

The Pan-STARRS1 Surveys (PS1) and the PS1 public science archive have been made possible through contributions by the Institute for Astronomy, the University of Hawaii, the Pan-STARRS Project Office, the Max-Planck Society and its participating institutes, the Max Planck Institute for Astronomy, Heidelberg and the Max Planck Institute for Extraterrestrial Physics, Garching, The Johns Hopkins University, Durham University, the University of Edinburgh, the Queen's University Belfast, the Harvard-Smithsonian Centre for Astrophysics, the Las Cumbres Observatory Global Telescope Network Incorporated, the National Central University of Taiwan, the Space Telescope Science Institute, the National Aeronautics and Space Administration under Grant No. NNX08AR22G issued through the Planetary Science Division of the NASA Science Mission Directorate, the National Science Foundation Grant No. AST-1238877, the University of Maryland, Eotvos Lorand University (ELTE), the Los Alamos National Laboratory, and the Gordon and Betty Moore Foundation. \\

This publication makes use of data products from the Two Micron All Sky Survey, which is a joint project of the University of Massachusetts and the Infrared Processing and Analysis Centre/California Institute of Technology, funded by the National Aeronautics and Space Administration and the National Science Foundation. \\

This work has made use of data from the European Space Agency (ESA) mission {\it Gaia} (\url{https://www.cosmos.esa.int/gaia}), processed by the {\it Gaia} Data Processing and Analysis Consortium (DPAC, \url{https://www.cosmos.esa.int/web/gaia/dpac/consortium}). Funding for the DPAC has been provided by national institutions, in particular the institutions participating in the {\it Gaia} Multilateral Agreement. \\ 

This work has made use of the following softwares: \texttt{astropy} \citep{Astropy_2018}, \texttt{dustmaps} \citep{dustmaps2018}, \texttt{plotly} \citep{plotly}, \texttt{healpy} \citep{Healpy_2005}, \texttt{dynesty} \citep{Speagle2019}, and \texttt{glue} \citep{glue2015}. \\

\section*{Data Availability}

The data underlying this article are available on the Harvard
Dataverse, at \href{https://dataverse.harvard.edu/dataverse/NPS}{https://dataverse.harvard.edu/dataverse/NPS}.




\bibliographystyle{mnras}
\bibliography{reference1}



\appendix

\section{Dynesty Setup Used}\label{A}

To sample for our set of parameters using the \dynesty\ code, we use the following setup: \\

\begin{lstlisting}

sampler=dynesty.NestedSampler (log-likelihood,
prior-transform, ndim, bound=`multi',
sample=`rwalk', update_interval=6., Ndraws=20., 
nlive=300,  walks=25)
    sampler.run_nested (dlogz=0.1)
\end{lstlisting}





\bsp	
\label{lastpage}
\end{document}